%% file: camera-ready.tex
\documentclass[sigconf]{acmart}

\AtBeginDocument{%
  \providecommand\BibTeX{{%
    \normalfont B\kern-0.5em{\scshape i\kern-0.25em b}\kern-0.8em\TeX}}}
    
\copyrightyear{2024}
\acmYear{2024}
\setcopyright{acmlicensed}\acmConference[MRAC '24]{Proceedings of the 2nd International Workshop on Multimodal and Responsible Affective Computing}{October 28-November 1 2024}{Melbourne, VIC, Australia}
\acmBooktitle{Proceedings of the 2nd International Workshop on Multimodal and Responsible Affective Computing (MRAC '24), October 28-November 1 2024, Melbourne, VIC, Australia}
\acmDOI{10.1145/3689092.3689404}
\acmISBN{979-8-4007-1203-6/24/10}

\usepackage[utf8]{inputenc} 
\usepackage[T1]{fontenc}    

\usepackage{multirow}

\usepackage{url}            
\usepackage{booktabs}       
\usepackage{amsfonts}       
\usepackage{nicefrac}       
\usepackage{microtype}      
\usepackage{algpseudocode}
\usepackage{bm}
\usepackage{mathtools}
\usepackage{algorithm}
\usepackage{algpseudocode}
\usepackage{lipsum} 
\usepackage{placeins}

\begin{document}

\title[SZTU-CMU: Improving Emotion-LLaMA for MER2024]{SZTU-CMU at MER2024: Improving Emotion-LLaMA with Conv-Attention for Multimodal Emotion Recognition}


\author{Zebang Cheng\textsuperscript{*}}
\affiliation{%
  \institution{ShenZhen Technology University}
  \city{Shenzhen}
  \country{China}}
\email{zebang.cheng@gmail.com}

\author{Shuyuan Tu\textsuperscript{*}}
\affiliation{%
  \institution{Carnegie Mellon University}
  \city{Pittsburgh}
  \country{USA}}
\email{francisshuyuan@gmail.com}

\author{Dawei Huang\textsuperscript{*}}
\affiliation{%
  \institution{ShenZhen Technology University}
  \city{Shenzhen}
  \country{China}}
\email{huangdawei2023@email.szu.edu.cn}

\author{Minghan Li}
\affiliation{%
  \institution{Carnegie Mellon University}
  \city{Pittsburgh}
  \country{USA}}
\email{lperlpm@gmail.com}

\author{Xiaojiang Peng\textsuperscript{\dag}}
\affiliation{%
  \institution{ShenZhen Technology University}
  \city{Shenzhen}
  \country{China}}
\email{pengxiaojiang@sztu.edu.cn}

\author{Zhi-Qi Cheng\textsuperscript{\dag}}
\affiliation{%
  \institution{Carnegie Mellon University}
  \city{Pittsburgh}
  \country{USA}}
\email{zhiqic@cs.cmu.edu}

\author{Alexander G. Hauptmann}
\affiliation{%
  \institution{Carnegie Mellon University}
  \city{Pittsburgh}
  \country{USA}}
\email{Alex@cs.cmu.edu}


\thanks{Z. Cheng (Emotion-LLaMA), S. Tu (Conv-Attention), D. Huang (feature engineering) contrib. equally. M. Li (replicated last year's champ solution).}

\thanks{X. Peng \& Z-Q. Cheng (corresponding authors, guided research/system, organized and rewrote the paper). A. Hauptmann (valuable insights \& suggestions).}

\renewcommand{\shortauthors}{Zebang Cheng et al.}


\begin{abstract}
This paper presents our winning approach for the \textit{MER-NOISE} and \textit{MER-OV} tracks of the \textit{MER2024 Challenge} on multimodal emotion recognition. Our system leverages the advanced emotional understanding capabilities of \textit{Emotion-LLaMA} to generate high-quality annotations for unlabeled samples, addressing the challenge of limited labeled data. To enhance multimodal fusion while mitigating modality-specific noise, we introduce \textit{Conv-Attention}, a lightweight and efficient hybrid framework. Extensive experimentation validates the effectiveness of our approach. In the \textit{MER-NOISE} track, our system achieves a state-of-the-art weighted average F-score of 85.30\%, surpassing the second and third-place teams by 1.47\% and 1.65\%, respectively. For the \textit{MER-OV} track, our utilization of \textit{Emotion-LLaMA} for open-vocabulary annotation yields an 8.52\% improvement in average accuracy and recall compared to \textit{GPT-4V}, securing the highest score among all participating large multimodal models. The code and model for \textit{Emotion-LLaMA} are available at \url{https://github.com/ZebangCheng/Emotion-LLaMA}.
\end{abstract}



\maketitle

\input{sections/01_introduction}

\input{sections/02_related_work}
\input{sections/03_method}
\input{sections/04_experiment}

\bibliographystyle{ACM-Reference-Format}
\bibliography{camera-ready}

\end{document}

%% file: sections/01_introduction.tex
\section{Introduction}

\begin{figure*}[t]
\centering
\includegraphics[width=0.9\linewidth]{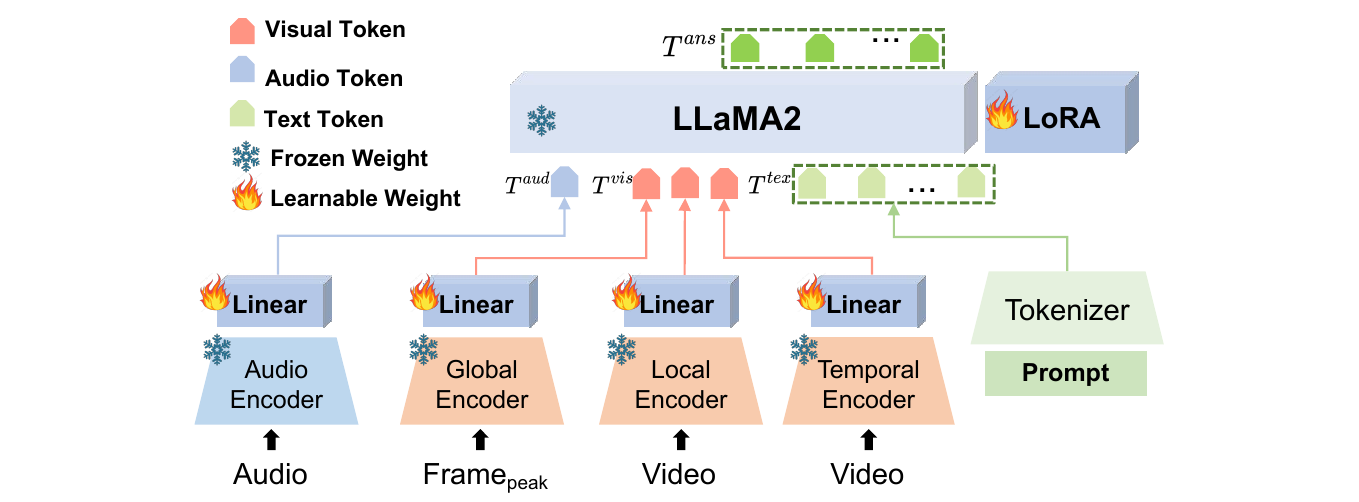}
\vspace{-1.0em}
\caption{Overview of the Emotion-LLaMA architecture, which integrates audio, visual, and text inputs for advanced multimodal emotion recognition and reasoning. The model aligns and fuses audio and visual features into a shared semantic space, thereby enhancing the contextual understanding of textual inputs. Emotion-LLaMA leverages multiple visual encoders to capture global, local, and temporal visual aspects, which are then combined with audio and text features to generate detailed emotion descriptions. For further details, refer to the original Emotion-LLaMA paper~\cite{Emotion-LLaMA}.}
\vspace{-0.5em}
\label{emotion-llama_framework}
\end{figure*}

\textit{Multimodal Emotion Recognition (MER)} aims to integrate information from various modalities—such as text, speech, and visual cues—to accurately identify and understand human emotions. This field has shown great promise in applications ranging from human-computer interaction to mental health care and education. However, achieving robust performance in real-world scenarios remains a significant challenge. The \textit{MER2024 Challenge} addresses these challenges through two specialized tracks: \textit{MER-NOISE} and \textit{MER-OV}.

The \textit{MER-NOISE} track focuses on enhancing noise robustness in emotion recognition systems. In practical settings, noise is pervasive, making it difficult to ensure that audio streams are free of distortions and video frames maintain high resolution. This track targets the two most prevalent types of noise: audio additive noise and image blur. Participants are encouraged to employ data augmentation techniques~\cite{pise2008survey} and other innovative methods~\cite{liu2022umt} to improve the resilience of emotion recognition systems against these noise factors.

The \textit{MER-OV} track introduces the concept of open-vocabulary emotion recognition, addressing the inherent subjectivity and ambiguity in emotion labeling. Traditional datasets often constrain label spaces to a few discrete categories, relying on multiple annotators and majority voting to determine the most likely label. This approach can overlook correct but non-candidate or minority labels, leading to potential inaccuracies. The \textit{MER-OV} track challenges participants to generate any number of labels across diverse categories, striving for a more nuanced and precise description of emotional states~\cite{MER2024}.

To address these challenges, we propose a robust system that integrates the advanced capabilities of \textit{Emotion-LLaMA} for generating high-quality labels with a \textit{Conv-Attention} model designed for efficient multimodal feature fusion. Our approach is detailed in Figure~\ref{framework}, where we outline the workflow and demonstrate how each component contributes to overcoming the limitations of existing methods.
A key limitation of previous approaches~\cite{cheng2023semi} lies in their reliance on models to generate pseudo-labels for unlabeled data, which are then used to augment training datasets. The effectiveness of this strategy depends heavily on the initial model's quality—if the model lacks robustness, it can produce low-quality pseudo-labels, which may introduce errors in subsequent training phases. To mitigate this issue, we introduce \textit{Emotion-LLaMA}~\cite{Emotion-LLaMA}, a model specifically designed to generate high-quality pseudo-labels for the unlabeled samples in the \textit{MER2024} dataset. As illustrated in Figure~\ref{emotion-llama_framework}, \textit{Emotion-LLaMA} processes inputs from multiple modalities, utilizing visual and auditory features as contextual information to enhance the interpretation of text-based emotions. This approach ensures robust multimodal emotion understanding, even in the presence of loss or noise of modality.

\begin{table*}[t]
    \centering
    \caption{Prompts for generating emotion-related descriptions using Large Multimodal  Models (LMMs). Prompts marked with $^\dag$ output only emotion categories, while those with $^\ddag$ provide complete emotion descriptions. These prompts direct the models to focus on key aspects of the input, ensuring high-quality, contextually rich outputs.}
    \vspace{-0.5em}
    \label{tab:prompt}
    \begin{tabular}{p{3.5cm}|p{2cm}|p{10cm}}
        \hline
        Models & Language & Prompt \\
        \hline
        \multirow{2}{*}{Emotion-LLaMA$^\dag$} & \multirow{2}{*}{English} & Please determine which \textcolor[rgb]{0.93,0.0,0.47}{emotion label} in the video represents: happy, sad, neutral, angry, worried, surprise. \\

        \hline
        \multirow{2}{*}{Emotion-LLaMA$^\ddag$} & \multirow{2}{*}{English} & Please analyze all the \textcolor[rgb]{0.93,0.0,0.47}{clues} in the video and \textcolor[rgb]{0.93,0.0,0.47}{reason out} the emotional label of the person in the video. \\
        
        \hline
        \multirow{4}{*}{LLaMA-3} & \multirow{4}{*}{English} & You are an emotion analysis expert. Please analyze the input multimodal emotion description and \textcolor[rgb]{0.93,0.0,0.47}{output keywords} related to the emotion description. \newline
        Input: [Multimodal Emotion Description] \newline
        Output: \\
        
        \hline
        \multirow{3}{*}{Qwen1.5-32B} & \multirow{3}{*}{Chinese} & Please \textcolor[rgb]{0.93,0.0,0.47}{analyze} the provided text content and \textcolor[rgb]{0.93,0.0,0.47}{classify emotions} into six categories: [neutral, angry, happy, sad, worried, surprise], and \textcolor[rgb]{0.93,0.0,0.47}{explain} the specific reasons: <Text> \\

        \hline
        \multirow{3}{*}{Baichuan-13B~(prompt 1)}  & \multirow{3}{*}{Chinese} & Please \textcolor[rgb]{0.93,0.0,0.47}{analyze} the provided text content and \textcolor[rgb]{0.93,0.0,0.47}{classify emotions} into six categories: [neutral, angry, happy, sad, worried, surprise], and \textcolor[rgb]{0.93,0.0,0.47}{explain} the specific reasons: <Text> \\

        \hline
        \multirow{2}{*}{Baichuan-13B~(prompt 2)} & \multirow{2}{*}{Chinese} & Please \textcolor[rgb]{0.93,0.0,0.47}{analyze} the provided text content and \textcolor[rgb]{0.93,0.0,0.47}{classify emotions} into six categories: [neutral, angry, happy, sad, worried, surprise]: <Text> \\
        
        \hline
        \multirow{1}{*}{Baichuan-13B~(prompt 3)} & \multirow{1}{*}{Chinese} & Please \textcolor[rgb]{0.93,0.0,0.47}{analyze} the provided text content: <Text> \\
        
        \hline
    \end{tabular}
\end{table*}

In the feature extraction stage, we leverage high-performing unimodal models referenced in the official baseline papers~\cite{MER2024,MERBench2023}, such as \textit{HuBERT}~\cite{Hubert} and \textit{CLIP}~\cite{CLIP}. Our experiments revealed that visual modality models are particularly vulnerable to noise, prompting us to pre-train \textit{MAE}~\cite{MAE} and \textit{VideoMAE}~\cite{VideoMAE} on the unlabeled samples from the \textit{MER2024} dataset. This pre-training effectively captures both static and dynamic visual expression features. Additionally, to enhance the accuracy of textual feature extraction, we employed prompt-based strategies for models like \textit{Qwen}~\cite{Qwen, qwen1.5} and \textit{Baichuan}~\cite{yang2023baichuan}, which were carefully evaluated for their effectiveness in capturing emotion-related information.

Despite the strong performance of \textit{Emotion-LLaMA} in \textit{MER}, its substantial computational overhead and slow iteration cycle present challenges. To address these, we propose \textit{Conv-Attention}, a lightweight and efficient hybrid framework that combines convolutional and global attention mechanisms for feature fusion. \textit{Conv-Attention} leverages the inductive biases inherent in convolutional operations, allowing the model to perform effectively even with limited data. By integrating a simple attention mechanism with multiple convolutional blocks, the model can prioritize critical features while minimizing the impact of noise. The attention mechanism excels in querying features from a global perspective, while the convolutional operation focuses on capturing fine-grained semantic details within a limited receptive field. This combined approach mitigates the disadvantages of each individual method, enhancing overall model performance.

In summary, our team makes the following contributions:
\begin{itemize}
\item[$\bullet$] In the \textit{MER-OV} track, we utilize \textit{Emotion-LLaMA} to extract multimodal emotion descriptions and employ \textit{LLaMA-3} for open-vocabulary annotation. The results from \textit{Emotion-LLaMA} serve as high-quality annotations for the unlabeled samples in the \textit{MER2024} dataset, addressing the need for extensive and accurate training data.
\item[$\bullet$] In the \textit{MER-NOISE} track, we introduce \textit{Conv-Attention}, a lightweight and effective hybrid framework for feature fusion that combines convolution and global attention mechanisms. Our model achieves comprehensive feature fusion by integrating the limited receptive fields of convolutions with the global querying capabilities of attention.
\item[$\bullet$] Our approach achieves a state-of-the-art weighted average F-score of 85.30\% in the \textit{MER-NOISE} track, outperforming the second and third-place entries by 1.47\% and 1.65\%, respectively. In the \textit{MER-OV} track, the application of \textit{Emotion-LLaMA} for open-vocabulary annotation resulted in an 8.52\% improvement in average accuracy and recall compared to \textit{GPT-4V}, significantly enhancing our competitive position.
\end{itemize}

%% file: sections/02_related_work.tex
\section{Related Work}

\subsection{Multimodal Emotion Recognition}
\textit{Multimodal Emotion Recognition (MER)} has become a focal point in multimedia research, driven by the limitations of single-modal approaches in handling noise and ensuring robustness. The rise of multimedia sensors has shifted research towards multimodal datasets from real-world scenarios, emphasizing the importance of feature alignment and fusion~\cite{lian2023mer, jiang2020dfew, aguilera2023assessment, poria2018meld}. Early \textit{MER} approaches utilized separate models for feature extraction across different modalities, such as ResNet~\cite{he2016resnet}, MAE~\cite{MAE}, VideoMAE~\cite{VideoMAE}, BERT~\cite{BERT}, and HuBERT~\cite{Hubert}, followed by basic linear fusion layers~\cite{lian2023mer, cheng2023semi, zhou2019exploring}. However, these simplistic models struggled to capture the complexity of multimodal data, prompting the development of more sophisticated cross-attention-based fusion models~\cite{ding2023stable,ding2022letr,chen2021crossvit}. Despite their advancements, these fusion techniques~\cite{sun2023efficient,liu2022umt,nagrani2021attention,li2024uni} often lead to competition between modalities, where dominant modalities disproportionately influence the results.

To address these challenges, \textit{PMR}~\cite{lv2021progressive} introduced a common message hub to better capture cross-modal dynamics. Subsequent research has focused on pre-fusion alignment, as seen in \textit{EmotionCLIP}~\cite{zhang2023learning}, which aligns temporal visual and textual data, and \textit{VAT}~\cite{ding2023learning}, which aligns visual with audio features. However, these models still face challenges, including the need for large datasets and the difficulty of capturing fine-grained emotional features due to a reliance on global attention mechanisms.

Our work overcomes these limitations by leveraging the advanced capabilities of \textit{Emotion-LLaMA} for generating high-quality pseudo-labels and introducing a \textit{Conv-Attention} model for efficient multimodal feature fusion, significantly improving the robustness and accuracy of emotion recognition in noisy environments.

\subsection{Large Models in Emotion Understanding}
The advent of large multimodal models (\textit{LMMs}) has revolutionized emotion understanding, providing unprecedented inferential capabilities~\cite{alayrac2022flamingo, shikra, peng2023kosmos, wang2023visionllm, FromLLM2LMM}. Instruction-tuning techniques, pioneered by models such as \textit{InstructionGPT}~\cite{ouyang2022instruct-tuning}, \textit{FLAN}~\cite{chung2022scaling}, and \textit{OPT-IML}~\cite{iyer2022opt-iml}, have further expanded the practical applications of these models~\cite{wang2022self_instruct, wang2022benchmarking}. In the context of emotion recognition, \textit{InstructERC}~\cite{lei2023instructerc} has advanced conversation-based emotion recognition by introducing additional emotion alignment tasks. \textit{DFER-CLIP}~\cite{zhao2023prompting}, built on the \textit{CLIP} model~\cite{CLIP}, has shown promise in dynamic facial expression recognition. \textit{MER-MCE}~\cite{cheng2024mips} has pushed the boundaries by inferring the causes of emotional triggers in conversations through multimodal inputs. Notably, \textit{GPT-4V}~\cite{lian2024gpt} has demonstrated strong capabilities in generalized emotion recognition tasks.

Despite these advances, most approaches rely on single emotion labels, often neglecting non-candidate or minority yet correct labels. Addressing the need for more nuanced emotion descriptions in real-world contexts, \textit{AffectGPT}~\cite{lian2023explainable} proposes a multimodal, explainable, open-vocabulary emotion recognition approach, using \textit{GPT-4V} to generate visual and acoustic signals and extract reliable labels. \textit{EmoVIT}~\cite{xie2024emovit} further contributes by generating visual emotion instruction data using paired annotations.

Building on these developments, our approach utilizes \textit{Emotion-LLaMA}~\cite{Emotion-LLaMA} to generate detailed multimodal emotional descriptions, resulting in comprehensive open-vocabulary labels. \textit{Emotion-LLaMA's} capability to align multimodal features within a semantic space allows it to maintain robust emotion understanding even in the presence of modality loss or noise, significantly enhancing the accuracy and robustness of emotion recognition systems in real-world applications.

%% file: sections/03_method.tex
\begin{figure*}[t]
    \centering
    \includegraphics[width=\linewidth]{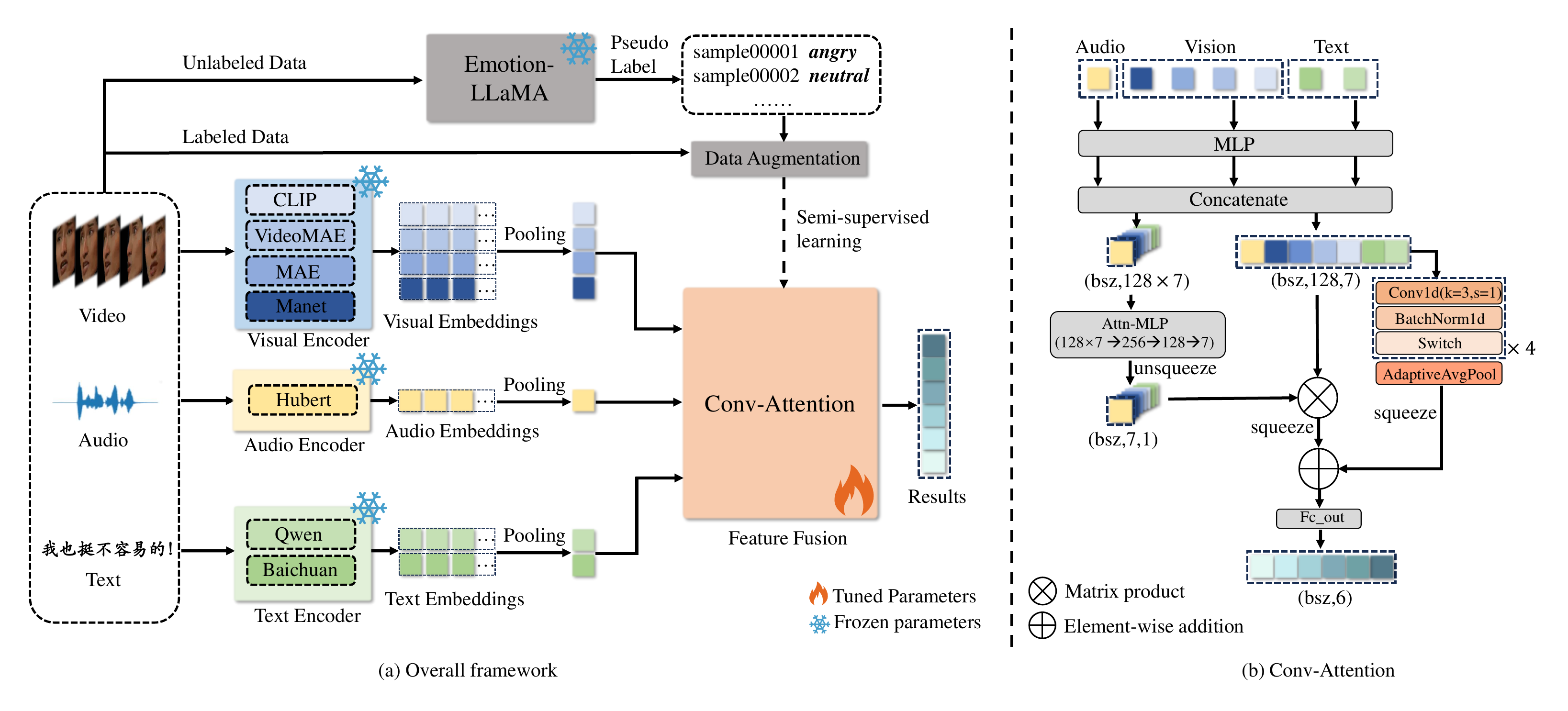}
    \vspace{-2em}
    \caption{Overview of our framework for MER2024. In the feature extraction phase, frozen encoders extract features from text, video, and audio, which are pooled to integrate multimodal information. In the feature fusion stage, our Conv-Attention mechanism is applied, as detailed in part (b) of the figure. The pre-trained Emotion-LLaMA~\cite{Emotion-LLaMA} model generates pseudo-labels, which are combined with original labeled data, enhancing the dataset through augmentation. Finally, the augmented dataset is used to train the Conv-Attention model, boosting the performance and robustness of our emotion recognition system.}
    \vspace{-0.5em}
    \label{framework}
\end{figure*}

\section{Methodology}
This section presents our method that secured the highest performance in Track 2: MER-NOISE at the MER 2024 contest. First, we detail our feature extraction process~(Sec.~\ref{sec:feature-extraction}). Next, we describe how \textit{Emotion-LLaMA} generates multimodal emotional descriptions and derives high-quality emotion labels~(Sec.~\ref{sec:emotion-llama}). Finally, we explain the Conv-Attention model used for feature fusion~(Sec.~\ref{sec:conv-attention}).

\subsection{Multimodal Feature Engineering}\label{sec:feature-extraction}
We employed domain-specific models to extract unimodal features from auditory, visual, and textual data, with each model leveraging prior knowledge tailored to its respective domain.

\subsubsection{Auditory Modality}
The MER2024 dataset contains exclusively Mandarin speech dialogues, prompting the selection of audio encoders that support the Chinese language. We prioritized Chinese-Hubert \cite{Hubert} and emotion2vec \cite{ma2023emotion2vec}, with Chinese-Hubert being a variant of Hubert pre-trained on Chinese datasets. This model excels in processing Mandarin, producing high-quality embeddings suitable for complex emotion detection, making it ideal for the challenges presented in MER2024.
To enhance the robustness and accuracy of our auditory emotion recognition pipeline, we also incorporated multilingual models such as Whisper \cite{Whisper}, VGGish \cite{hershey2017cnn}, and eGeMAPS \cite{eyben2016geneva}. Whisper, a significant advancement in Automatic Speech Recognition (ASR), combined with VGGish and eGeMAPS, provided a comprehensive audio processing framework.

\subsubsection{Visual Modality}
Building on the experience and results from the MER2023 competition, we selected a series of high-performing visual encoders for comparative analysis, including CLIP \cite{CLIP}, MAE \cite{MAE}, VideoMAE \cite{VideoMAE}, and MANet \cite{MANet}.
CLIP, pre-trained on large-scale image-text pairs, excels at associating images with textual descriptions, making it particularly effective for affective computing tasks. MAE, designed for the visual domain, reconstructs obscured segments of input images, compelling the model to learn both global and local features, which is advantageous for facial recognition and emotion analysis.
VideoMAE extends MAE's concept to video by applying random masking to video frames and training the model to reconstruct missing parts, effectively leveraging spatiotemporal characteristics for tasks such as video classification. Both MAE and VideoMAE were further fine-tuned to enhance their performance in noisy environments.

\subsubsection{Textual Modality}
Given that the subtitles extracted from audio are primarily in Chinese, we focused on models with strong Chinese language proficiency, including ChatGLM2 \cite{du2022glm}, Qwen \cite{Qwen}, and Baichuan \cite{yang2023baichuan}. These models, pre-trained on extensive Chinese corpora, are particularly effective for handling Chinese text inputs.
We also utilized multilingual models such as RoBERTa \cite{liu2019roberta}, MacBERT \cite{cui2020revisiting}, and BLOOM \cite{workshop2022bloom}. Inspired by In-Context Learning \cite{GPT-3} and Chain-of-Thought (CoT) \cite{CoT} techniques, we employed a prompt strategy to enhance feature extraction. This approach involved generating emotion-associated descriptions by appending a designed prompt before the text input, as formalized in the following equation:
\begin{equation}
\label{eq:prompt}
\begin{aligned}
&T' = \text{prompt} \oplus T \\
&T_{\text{embedding}} = M(T')
\end{aligned}
\end{equation}
Here, \(T'\) represents the augmented text input, \(\oplus\) denotes the concatenation operation, and \(M\) is the language model used for feature extraction. This prompt-based method guides the model to capture emotional cues more effectively, resulting in richer and more accurate emotional descriptions.

\subsection{Emotion-LLaMA Pseudo-Labeling}\label{sec:emotion-llama}
\textit{Emotion-LLaMA}~\cite{Emotion-LLaMA}, developed in our previous work, is a multimodal emotion recognition model that supports inputs across text, audio, and visual domains. By aligning audio and visual features within a shared semantic space as contextual information for the text modality, Emotion-LLaMA excels in multimodal emotion recognition and reasoning tasks. We leveraged Emotion-LLaMA's capabilities for the MER2024 competition.

To address the challenge of limited labeled data, especially for the MER-NOISE track, we used Emotion-LLaMA to generate pseudo-labels. By performing multimodal emotion recognition on 20,000 unlabeled samples, we significantly augmented the training set with pseudo-labeled data. This approach not only increased the volume of training data but also introduced greater diversity, thereby enhancing the model's generalization capabilities.

\subsubsection{Prompt Design and Data Processing}
We designed specific prompts for Emotion-LLaMA and LLaMA-3 to extract detailed emotion-related descriptions and labels. These prompts guide the models to focus on relevant aspects of the input data, improving the quality and relevance of the generated labels. Table~\ref{tab:prompt} provides examples of these prompts, illustrating their effectiveness in eliciting precise and informative responses.

We processed the data by feeding it into Emotion-LLaMA with a simple instruction prompt to obtain emotion-related descriptions and category labels, as formalized below:
\begin{equation}
\label{eq:input+prompt}
\begin{aligned}
\hat{\mathcal{T}}_{description}=E( \mathcal{A}'_u , \mathcal{V}'_u , \mathcal{T}'_u ; \mathcal{P ^\ddag}) \\
\hat{\mathcal{L}}=E( \mathcal{A}'_u , \mathcal{V}'_u , \mathcal{T}'_u ; \mathcal{P ^\dag})
\end{aligned}
\end{equation}
where $E$ represents Emotion-LLaMA, $\hat{\mathcal{T}}_{description}$ is the emotion-related description generated for the MER-OV track, $\hat{\mathcal{L}}$ denotes the pseudo-label set, and $\mathcal{A}'_u , \mathcal{V}'_u , \mathcal{T}'_u$ are the data in the audio, visual, and textual modalities, respectively. The prompts $\mathcal{P}^\dag$ and $\mathcal{P}^\ddag$ are used for multimodal emotion classification and reasoning, respectively.

\subsubsection{Keyword Extraction and Dataset Augmentation}
We employed LLaMA-3 as a keyword extractor to convert emotional descriptions into labels, which were then used as the final prediction results for MER-OV. These pseudo-labeled samples were combined with the Train\&Val dataset from MER2024 to create the training set for the multimodal fusion model:
\begin{equation}
\label{eq:text2labels}
\begin{aligned}
&\mathcal{L}_{OV}=LLaMA(\hat{\mathcal{T}}_{description})\\
&\mathcal{D}_{aug}=\mathcal{D}_l \cup (\mathcal{D}'_u, \hat{\mathcal{L}})
\end{aligned}
\end{equation}
where $\mathcal{L}_{OV}$ represents the open vocabulary labels, $\mathcal{D}_l$ is the labeled dataset from MER2024, $\mathcal{D}'_u$ is the unlabeled dataset, and $\mathcal{D}_{aug}$ is the augmented dataset. By leveraging the capabilities of Emotion-LLaMA, our methodology significantly enhances the volume and quality of data available for training, effectively addressing the issue of sample scarcity.

\subsection{Multimodal Feature Fusion}\label{sec:conv-attention}
To address the limitations of pure attention mechanisms, we employed the Conv-Attention structure, as illustrated in Figure \ref{framework}(b). This structure integrates convolutional blocks with attention mechanisms to introduce inductive biases, improving performance on limited-scale data. The convolutional branch captures semantic details due to its limited receptive fields, while the attention branch focuses on global, emotionally salient features.

We began by using a multilayer perceptron (MLP) to standardize the channel depth of features from different modalities (audio, visual, text) to a common dimension. Each feature is represented as $(\text{batch}, \text{depth})$. These features were then concatenated along their embedding depths and sequence lengths to obtain hybrid features $\bm{F}_{d}$ and $\bm{F}_{s}$:
\begin{equation}
\label{eq:mlp_dimension}
\begin{aligned}
     &\bm{F}_{d} = \mathtt{Concat}(\mathtt{MLP}(\bm{A}), \mathtt{MLP}(\bm{V}), \mathtt{MLP}(\bm{T}), \text{dim}=1), \\
     &\bm{F}_{s} = \mathtt{Stack}(\mathtt{MLP}(\bm{A}), \mathtt{MLP}(\bm{V}), \mathtt{MLP}(\bm{T}), \text{dim}=2)
\end{aligned}
\end{equation}
where $\bm{A}=\left \{ \bm{A}_{hubert} \right \}$, $\bm{V}=\left \{ \bm{V}_{clip},\bm{V}_{videomae},\bm{V}_{mae},\bm{V}_{manet} \right \}$, and $\bm{T}=\left \{ \bm{T}_{qwen},\bm{T}_{baichuan} \right \}$ refer to audio, visual, and text features, respectively.

In the attention branch, we applied $\mathtt{Attn\_MLP}(\cdot)$ to $\bm{F}_{d}$, downsampling its embedding depth to align with the sequence length of $\bm{F}_{s}$. We then performed a matrix product between the downsampled $\bm{F}_{d}$ and $\bm{F}_{s}$ to obtain the attention-based fusion feature $\bm{F}_{attn}$, which enhances the model's ability to identify emotionally salient components across different modalities:
\begin{equation}
\label{eq:attention}
\begin{aligned}
     \bm{F}_{attn}=\mathtt{Unsqueeze}(\mathtt{Attn\_MLP}(\bm{F}_{d}), \text{dim}=-1) \times \bm{F}_{s}
\end{aligned}
\end{equation}
where $\times$ denotes the matrix product, and $\mathtt{Unsqueeze}(\cdot)$ denotes the unsqueeze operation.

In the convolutional branch, we designed a lightweight convolution block consisting of a convolution layer $\mathtt{Conv1d}(\cdot)$, batch normalization $\mathtt{BN}(\cdot)$, and an activation function $\mathtt{Swish}(\cdot)$. The convolutional branch includes $\bm{N}$ convolution blocks and an adaptive average pooling layer $\mathtt{Pool}(\cdot)$, which reshapes the ultimate fusion features. The convolutional operators’ inductive bias and limited receptive fields encourage the model to focus on fine-grained details, enhancing robustness, particularly when trained on limited-scale datasets:
\begin{equation}
\label{eq:convolution}
\begin{aligned}
    & \bm{F}_{conv}^{k} = \mathtt{Swish}(\mathtt{BN}(\mathtt{Conv1d}(\bm{F}_{conv}^{k-1}))), \\
    & \bm{F}_{conv} = \mathtt{Pool}(\bm{F}_{conv}^{N})
\end{aligned}
\end{equation}
where $k (k=1,2,...,N)$ refers to the index of the convolution block, and $\bm{F}_{conv}$ indicates the final convolution-based fusion feature. Note that $\bm{F}_{conv}^{0}=\bm{F}_{s}$.

Finally, we employed a residual connection to combine $\bm{F}_{conv}$ and $\bm{F}_{attn}$ into the final fusion feature $\bm{F}_{fusion}$, which is then fed into a linear classification head $\mathtt{FC}_{out}(\cdot)$ for emotion prediction:
\begin{equation}
\label{eq:prediction}
\begin{aligned}
    & \bm{emotion}_{pred} = \mathtt{FC}_{out}(\bm{F}_{conv}+\bm{F}_{attn})
\end{aligned}
\end{equation}
The pseudo-labels generated by Emotion-LLaMA were integrated with the 5030 Train\&Val datasets from MER2024 to form the training set for our Conv-Attention model, as depicted in Figure \ref{framework}(b). This integration through data augmentation significantly improves the performance and robustness of our emotion recognition system.

%% file: sections/04_experiment.tex
\section{Experiments}
This section presents a comprehensive evaluation of our approach on Track 2 (MER-NOISE) and Track 3 (MER-OV) of the MER2024 competition. We begin by analyzing the performance of single-modal models, followed by multimodal fusion experiments, and conclude with ablation studies. Our goal is to provide detailed insights that can inform future research and practical applications in multimodal emotion recognition, particularly in noisy and open-vocabulary environments.

\subsection{Single-Modal Performance on Track 2: MER-NOISE}

We assessed several single-modal models on the MER-NOISE track to analyze the contributions of auditory, visual, and textual modalities to emotion recognition performance. The results are summarized in Table~\ref{tab:Unimodal_result_task2}.

\begin{table}[!ht]
    \centering
    \caption{Performance (\%) of Single-Modal Models on Track 2: MER-NOISE. $^*$: Using prompts to extract features as input for the large language model.}
    \label{tab:Unimodal_result_task2}
    \begin{tabular}{l|cc|c}
        \toprule
        \multirow{2}{*}{Features} & \multicolumn{2}{c|}{Train\&Val} & \multicolumn{1}{c}{MER-NOISE} \\
        & WAF $(\uparrow)$ & ACC $(\uparrow)$ & WAF $(\uparrow)$ \\
        \midrule
        \multicolumn{4}{c}{Audio Modality} \\
        \midrule
        eGeMAPS \cite{eyben2016geneva}              & 39.68 & 42.88 & 28.92 \\
        VGGish \cite{hershey2017cnn}                & 48.60 & 50.20 & 40.70 \\
        Whisper-base \cite{Whisper}                 & 56.65 & 57.08 & 41.26 \\
        emotion2vec \cite{ma2023emotion2vec}        & 56.08 & 56.48 & 45.66 \\
        HuBERT-large \cite{Hubert}                  & \textbf{72.77} & \textbf{72.96} & \textbf{72.67} \\
        \midrule
        \multicolumn{4}{c}{Visual Modality} \\
        \midrule
        MANet-RAFDB \cite{MANet}                    & 60.31 & 61.33 & 54.62 \\
        MAE-MER2024 \cite{MAE}                      & 61.48 & 62.40 & 51.11 \\
        VideoMAE \cite{VideoMAE}                    & 57.40 & 58.10 & 49.18 \\
        VideoMAE-MER2024 \cite{VideoMAE}            & 64.87 & 65.46 & 57.87 \\
        CLIP-large \cite{CLIP}                      & \textbf{66.73} & \textbf{67.28} & \textbf{58.80} \\
        \midrule
        \multicolumn{4}{c}{Text Modality} \\
        \midrule
        RoBERTa-large \cite{liu2019roberta}         & 52.66 & 52.92 & 49.06 \\
        ChatGLM2-6B \cite{du2022glm}                & 53.04 & 53.28 & 50.39 \\
        MacBERT-large \cite{cui2020revisiting}      & 52.19 & 52.47 & 50.24 \\
        BLOOM-7B \cite{workshop2022bloom}           & 53.13 & 53.30 & 50.38 \\
        Qwen1.5-32B \cite{qwen1.5}                  & 54.47 & 54.82 & 50.12 \\
        Qwen1.5-32B \cite{qwen1.5} $^*$             & \textbf{55.41} & \textbf{55.89} & 58.88 \\
        Baichuan-13B \cite{yang2023baichuan}        & 55.15 & 55.40 & 57.94 \\
        Baichuan-13B \cite{yang2023baichuan} $^*$   & 54.29 & 54.48 & \textbf{59.32} \\
        \bottomrule
    \end{tabular}
\end{table}

\subsubsection{Auditory Modality}
Given the limited availability of audio extraction models tailored for the Chinese language, we evaluated five models: eGeMAPS~\cite{eyben2016geneva}, VGGish~\cite{hershey2017cnn}, Whisper~\cite{Whisper}, emotion2vec~\cite{ma2023emotion2vec}, and Chinese-Hubert~\cite{Hubert}. The Chinese-Hubert model emerged as the top performer with a Weighted Average F-score (WAF) of 72.67\%. This superior performance can be attributed to its pre-training on Chinese datasets, which enables it to capture contextual representations more effectively than other models.

\subsubsection{Visual Modality}
In the visual modality, we evaluated four models: MANet~\cite{MANet}, MAE~\cite{MAE}, VideoMAE~\cite{VideoMAE}, and CLIP~\cite{CLIP}, along with versions fine-tuned for MER2024. CLIP achieved the highest WAF of 58.80\%, likely due to its extensive pre-training and multimodal learning capabilities. VideoMAE was further enhanced by domain-specific fine-tuning for emotion recognition tasks.

\subsubsection{Textual Modality}
For the textual modality, we focused on models with strong Chinese language proficiency. Baichuan-13B, when used with carefully designed prompts, attained the highest WAF of 59.32\%, closely followed by Qwen1.5-32B with a WAF of 58.88\%. The strong performance of these models can be attributed to their large-scale pre-training on Chinese corpora and the effective use of prompts, which significantly enhanced their ability to recognize and classify emotions in textual data. Further analysis of prompt designs, as shown in Table~\ref{tab:prompt_result_task2}, revealed that Prompt 1 provided the best performance across both single-modal and multimodal fusion scenarios.

\begin{table}[!ht]
    \centering
    \caption{Performance (\%) of different prompts on Track 2: MER-NOISE. The contents of the three prompts are detailed in Table~\ref{tab:prompt}. We select acoustic features from HuBERT (HB), visual features from CLIP (CL), and textual features from Baichuan (BC).}
    \label{tab:prompt_result_task2}
    \resizebox{0.98\linewidth}{!}{%
    \begin{tabular}{ccc|cc|c}
        \toprule
        \multicolumn{3}{c|}{Features} & \multicolumn{2}{c|}{Train\&Val} & \multicolumn{1}{c}{MER-NOISE} \\
        A & V & T & WAF $(\uparrow)$ & ACC $(\uparrow)$ & WAF $(\uparrow)$ \\
        \midrule
        - & -   & BC(w/o) &    55.15 & 55.40 & 57.94 \\
        - & -   & BC(w/ prompt 1) &    54.29 & 54.48 & 59.32 \\
        - & -   & BC(w/ prompt 2) &    57.87 & 61.15 & 58.66 \\
        - & -   & BC(w/ prompt 3) &    56.53 & 59.00 & 57.77 \\
        HB & CL & BC(w/o) &    78.58 & 79.96 & 78.73 \\
        HB & CL & BC(w/ prompt 1) &    \textbf{80.91}& \textbf{81.01} & \textbf{79.73}\\
        HB & CL & BC(w/ prompt 2) &    78.96& 79.33 & 78.71 \\
        HB & CL & BC(w/ prompt 3) &    79.11 &79.50 & 77.61 \\
        \bottomrule
    \end{tabular}
    }
\end{table}

\subsection{Multimodal Fusion on Track 2: MER-NOISE}
Leveraging the findings from the single-modal evaluations, we conducted multimodal fusion experiments by integrating features from the best-performing models in each modality. The results, as presented in Table~\ref{tab:Multimodal_result_task2}, demonstrate the effectiveness of our proposed Conv-Attention model. The configuration combining HuBERT, CLIP, VideoMAE, Qwen, and Baichuan features yielded the highest WAF and ACC scores of 81.59\% and 81.71\% on the Train\&Val dataset. On the MER-NOISE track, the optimal setup, which also included additional visual models (MAE and MANet), achieved a WAF of 80.10\%. These results underscore the effectiveness of multimodal fusion, particularly when employing our Conv-Attention model, which consistently outperformed other fusion strategies across all metrics (Table~\ref{tab:Multimodal_fusion_task2}) when trained on the augmented dataset.

\begin{table*}[!ht]
    \centering
    \caption{Performance (\%) of Multimodal Fusion Methods on MER-NOISE Track. $^*$ Denotes methods using prompt-extracted features as input for large language models.}
    \label{tab:Multimodal_result_task2}
    \begin{tabular}{ccccccc|cc|c}
        \toprule
        \multicolumn{7}{c|}{Features} & \multicolumn{2}{c|}{Train\&Val} & \multicolumn{1}{c}{MER-NOISE} \\
        A & V & V & V & V & T & T & WAF $(\uparrow)$ & ACC $(\uparrow)$ & WAF $(\uparrow)$ \\
        \midrule
        HuBERT & CLIP & - & - & - & Qwen $^*$ & -        & 78.43 & 78.67 & 77.44 \\
        HuBERT & CLIP & - & - & - & - & Baichuan $^*$   & 80.91 & 81.01 & 79.73 \\
        HuBERT & CLIP & - & - & - & Qwen $^*$ & Baichuan $^*$ & 80.50 & 80.63 & 79.79 \\
        HuBERT & CLIP & VideoMAE & - & - & Qwen $^*$ & -         & 79.22 & 79.23 & 77.10 \\
        HuBERT & CLIP & VideoMAE & - & - & - & Baichuan $^*$    & 80.44 & 80.48 & 78.49 \\
        HuBERT & CLIP & VideoMAE & - & - & Qwen $^*$ & Baichuan $^*$  & \textbf{81.59} & \textbf{81.71} & 79.63 \\
        HuBERT & CLIP & VideoMAE & MAE & - & Qwen $^*$ & -           & 78.98 & 78.93 & 76.80 \\
        HuBERT & CLIP & VideoMAE & MAE & - & - & Baichuan $^*$      & 81.52 & 81.59 & 79.03 \\
        HuBERT & CLIP & VideoMAE & MAE & - & Qwen $^*$ & Baichuan $^*$   & 81.49 & 81.55 & 79.93 \\
        HuBERT & CLIP & VideoMAE & MAE & MANet & Qwen $^*$ & -           & 79.35 & 79.48 & 77.30 \\
        HuBERT & CLIP & VideoMAE & MAE & MANet & - & Baichuan $^*$      & 81.45 & 81.57 & 79.93 \\
        HuBERT & CLIP & VideoMAE & MAE & MANet & Qwen $^*$ & Baichuan $^*$   & 81.40 & 81.53 & \textbf{80.10} \\
        \bottomrule
    \end{tabular}
\end{table*}

\begin{table}[!ht]
\caption{Performance (\%) Comparison of Multimodal Fusion Models on Track 2: MER-NOISE.}
\begin{center}
\renewcommand\arraystretch{1.1}
\scalebox{0.9}{
\begin{tabular}{lccc}
\toprule
Model                   & Train WAF       & Train ACC       & Noise WAF       \\ \midrule
MLP \cite{lian2023mer}                & 56.44 & 65.08 & 50.02 \\
Attention  \cite{MER2024}             & 82.29 & 82.59 & 83.48 \\
FBP \cite{zhou2019exploring}          & 80.98 & 81.17 & 83.21 \\
Convolution                           & 81.90  & 82.35 & 84.50  \\
Transformer                           & 80.81 & 81.23 & 84.55 \\
Conv-Attention (ours)          & \textbf{83.59} & \textbf{83.83} & \textbf{85.30} \\ \bottomrule
\end{tabular}
}
\end{center}
\label{tab:Multimodal_fusion_task2}
\end{table}

\subsection{Performance on Track 3: MER-OV}
For Track 3 (MER-OV), which addresses open-vocabulary emotion recognition, we evaluated various large language models using $\mbox{Accuracy}_s$, $\mbox{Recall}_s$, and their average (Avg). The results are detailed in Table~\ref{tab:OV}. Emotion-LLaMA outperformed other models in terms of average performance. The version that outputs only emotion categories achieved an accuracy of 83.43\%, while the version that generates complete emotion descriptions attained the highest recall of 62.59\% and an average score of 66.10\%. The success of Emotion-LLaMA can be attributed to three key factors: (1) emotion-specific pre-training on corpora rich in emotional content, enabling the model to capture subtle emotional nuances; (2) multi-task learning, which allows the model to excel in both emotion classification and description generation; and (3) an open-vocabulary design, which is well-suited for handling diverse and complex emotion descriptions. The trade-off observed between accuracy and recall suggests that while more detailed emotion descriptions enhance recall, they may also introduce a higher risk of misclassification.

\begin{table}[!ht]
  \caption{Performance (\%) of Single-Modal Models on Track 3: MER-OV. The ``avg'' column represents the average of ``$\mbox{Accuracy}_{\mbox{s}}$'' and ``$\mbox{Recall}_{\mbox{s}}$''. $^\dag$: Only outputs emotion categories; $^\ddag$: Outputs complete emotion descriptions.}
  \label{tab:OV}
  \begin{tabular}{lccc}
    \toprule
    Model & $\mbox{Accuracy}_{\mbox{s}}$ & $\mbox{Recall}_{\mbox{s}}$ & Avg \\
    \midrule
    Empty           & 0.00   & 0.00   & 0.00   \\
    Random          & 13.42  & 24.85  & 19.13  \\
    Ground Truth    & 93.37  & 52.51  & 72.94 \\
    \midrule    
    Valley \cite{luo2023valley}         & 20.16 & 13.26 & 16.71 \\
    Otter \cite{li2023otter}            & 29.64 & 23.04 & 26.34 \\
    PandaGPT \cite{su2023pandagpt}      & 35.75 & 31.57 & 33.66 \\
    Video-LLaMA \cite{zhang2023video}   & 31.08 & 32.26 & 31.67 \\
    VideoChat \cite{li2023videochat}    & 43.17 & 44.92 & 44.05 \\
    VideoChat2  \cite{li2024mvbench}    & 46.91 & 34.78 & 40.85 \\
    Video-ChatGPT \cite{maaz2023video}  & 46.20 & 39.33 & 42.77 \\
    SALMONN \cite{tang2023salmonn}      & 42.20 & 44.75 & 43.47 \\
    Qwen-Audio \cite{Qwen-audio}        & 55.12 & 32.91 & 44.02 \\
    mPLUG-Owl \cite{ye2023mplugowl}     & 44.80 & 46.54 & 45.67 \\
    AffectGPT \cite{lian2024affectgpt}  & 66.14 & 46.56 & 56.35 \\
    GPT-4V \cite{openai2023gpt4v}       & 56.19 & 58.97 & 57.58 \\
    Emotion-LLaMA \cite{Emotion-LLaMA} $^\dag$  & \textbf{83.43} & 47.49 & 65.46 \\
    Emotion-LLaMA \cite{Emotion-LLaMA} $^\ddag$ & 69.61 & \textbf{62.59} & \textbf{66.10} \\
    \bottomrule
  \end{tabular}
\end{table}

\subsection{Ablation Studies}
To gain deeper insights into the effectiveness of our proposed Conv-Attention model, we conducted a series of ablation studies to evaluate the impact of various components and hyperparameters.

\subsubsection{Component Ablation}
Table~\ref{table:ablation_components} presents the results of our component ablation study. The findings highlight several important observations: (1)~The ReLU activation function is crucial for introducing non-linearity, which significantly improves the model's ability to learn complex patterns. Removing ReLU resulted in a noticeable performance drop. (2)~The use of multiple convolutional blocks enhances the model's capability to capture hierarchical features, which is beneficial for recognizing multi-scale patterns in multimodal data. The best performance was observed with two convolutional blocks.(3)~Batch normalization plays a vital role in stabilizing the learning process and improving generalization. The inclusion of batch normalization layers led to a performance boost, likely due to their effect in reducing the internal covariate shift.

\begin{table}[!ht]
\caption{Performance Impact (\%) of Conv-Attention.}
\vspace{-2mm}
\begin{center}
\renewcommand\arraystretch{1.1}
\scalebox{0.9}{
\begin{tabular}{lccc}
\toprule
Model                   & Train WAF       & Train ACC       & Noise WAF       \\ \midrule
Relu                    & 80.52          & 80.69          & 81.70          \\
Attention(Conv-Block$\times$0)  & 81.20          & 81.43          & 81.48          \\
Conv-Block$\times$1             & 80.88          & 81.05          & 81.74          \\
Conv-Block$\times$2             & 81.00          & 81.09          & 82.09          \\
Conv-Block$\times$3             & 80.88          & 80.85          & 82.29          \\
\textit{w/o} Batch-Normalization & 80.96          & 81.13          & 81.64         \\ \midrule
Conv-Attention          & \textbf{81.37} & \textbf{81.45} & \textbf{82.68} \\ \bottomrule
\end{tabular}
}
\end{center}
\label{table:ablation_components}
\end{table}

\subsubsection{Impact of Data Ratio and Learning Rate}
We also explored the effects of varying data ratios and learning rates on model performance. Tables~\ref{table:ablation_ratio} and \ref{table:ablation_lr} summarize our findings. Increasing the data ratio consistently improved performance, with 100\% data usage yielding the highest scores across all metrics. This suggests that pseudo-labeling effectively augments the training data, enabling the model to learn from a larger and more diverse dataset. However, it is important to consider that while using 100\% of pseudo-labeled data can enhance performance, it may also introduce some noise. The optimal ratio may depend on the quality of the pseudo-labels.

\begin{table}[t]
    \centering
    \vspace{-2mm}
    \caption{Impact of Data Ratio on Performance(\%).}
    \vspace{-2mm}
    \label{table:ablation_ratio}
    \begin{tabular}{lccc}
        \hline
        \textbf{Data Ratio} & \textbf{Train WAF} & \textbf{Train ACC} & \textbf{Noise WAF} \\
        \hline
        20\% & 79.18 & 79.72 & 79.88 \\
        40\% & 80.06 & 80.36 & 81.29 \\
        60\% & 81.29 & 81.54 & 82.2 \\
        80\% & 81.98 & 82.19 & 82.5 \\
        100\% & \textbf{82.05} & \textbf{82.3} & \textbf{83.87} \\
        \hline
    \end{tabular}
\end{table}

\begin{table}[t!]
\vspace{-2mm}
\caption{Impact of Learning Rate on Performance(\%).}
\vspace{-2mm}
\begin{center}
\renewcommand\arraystretch{1}
\scalebox{1}{
\begin{tabular}{lccc}
\toprule
Learning Rate   & Train WAF   & Train ACC     & Noise WAF  \\ 
\midrule
1e-4     & \textbf{83.29} & \textbf{83.38} & 83.10 \\
5e-4     & 82.95 & 83.20 & 83.72 \\
1e-3     & 82.05 & 82.30 & \textbf{83.87} \\
5e-3     & 79.87 & 80.29 & 83.32  \\
1e-2     & 78.58 & 79.45 & 82.09 \\ 
5e-2     & 72.82 & 76.13 & 73.31  \\
\bottomrule
\end{tabular}
}
\end{center}
\label{table:ablation_lr}
\end{table}

Regarding learning rates, a rate of 1e-3 provided the best balance between convergence speed and model accuracy, achieving the highest Noise WAF. Lower learning rates (e.g., 1e-4) resulted in slower convergence, while higher rates (e.g., 5e-3 and above) caused unstable training and poor generalization, especially in noisy environments. These findings highlight the critical importance of proper hyperparameter tuning in achieving optimal performance, particularly in challenging multimodal and noisy settings.

\subsubsection{Modality Alignment}
Building on previous work~\cite{zong2023acformer}, we examined the impact of modality alignment on MER, with results presented in Table~\ref{tab:align}. Interesting phenomena emerged from these experiments:~(1)~Post-alignment, the scores of the previously weaker visual and textual modalities improved. However, this came at the cost of a performance decline in the best-performing audio modality, suggesting that alignment may sometimes dilute the complementary strengths of individual modalities.~(2)~The performance of the multimodal fusion dropped significantly after alignment. This indicates that while alignment may homogenize features across modalities, it can reduce the benefits derived from the diversity of information carried by different modalities.

\begin{table}[!ht]
    \centering
    \vspace{-2mm}
    \caption{Modality Alignment Impact on Performance (\%). Features: HuBERT (HB) for acoustic, CLIP (CL) for visual, Baichuan (BC) for textual.}
    \vspace{-2mm}
    \label{tab:align}
    \resizebox{0.95\linewidth}{!}{%
    \begin{tabular}{ccc|cc|c}
        \toprule
        \multicolumn{3}{c|}{Features} & \multicolumn{2}{c|}{Train\&Val} & \multicolumn{1}{c}{MER-NOISE} \\
        A & V & T & WAF $(\uparrow)$ & ACC $(\uparrow)$ & WAF $(\uparrow)$ \\
        \midrule
        HB              & -             & -                 &    72.77 & 72.96 & 72.67 \\
        HB(align)       & -             & -                 &    66.18 & 66.63 & 66.04 \\
        -               & CL          & -                 &    66.73 & 67.28 & 58.80 \\
        -               & CL(align)   & -                 &    69.07 & 69.50 & 65.93 \\
        -               & -             & BC          &    54.29 & 54.48 & 59.32 \\
        -               & -             & BC(align)   &    59.87 & 57.78 & 48.66 \\
        HB              & CL          & BC          &    80.91 & 81.01 & 79.73 \\
        HB(align)       & CL(align)   & BC(align)   &    73.01 & 73.29 & 70.67 \\
        \bottomrule
    \end{tabular}
    }
\end{table}

\subsection{Limitations \& Future Work}
Despite the effectiveness of our methods, several limitations remain that warrant further investigation: (1) Our research is limited to Chinese language data, highlighting the need for validation across other languages and cross-lingual scenarios. (2) The models are not optimized for real-time emotion recognition, indicating a need for improvements in computational efficiency. (3) While our ablation studies provide valuable insights, enhancing the interpretability of multimodal fusion models is crucial, particularly in understanding the contributions of each modality. (4) Our emphasis on the \textit{MER-NOISE} track underscores the importance of exploring model robustness across different noise types and levels.
Addressing these limitations is crucial for the continued advancement of \textit{multimodal emotion recognition} and the development of more flexible and effective systems for practical applications.

\section{Conclusion}
In this paper, we presented our winning approach for enhancing \textit{multimodal emotion recognition} in the \textit{MER2024 Challenge}, specifically targeting the \textit{MER-NOISE} and \textit{MER-OV} tracks. By leveraging the advanced capabilities of \textit{Emotion-LLaMA} to generate high-quality pseudo-labels and introducing a \textit{Conv-Attention} mechanism for efficient feature fusion, we significantly improved the robustness and accuracy of emotion recognition. Our method delivered state-of-the-art performance in the \textit{MER-NOISE} track with a weighted average F-score of 85.30\% and achieved top results in the \textit{MER-OV} track, enhancing average accuracy and recall by 8.52\% compared to GPT-4V. The integration of \textit{Emotion-LLaMA} was pivotal in achieving these results, underscoring its potential to advance the field of multimodal emotion recognition.

\section*{Acknowledgments}
The SZTU team acknowledges support from the National Natural Science Foundation of China (62176165), the Stable Support Projects for Shenzhen Higher Education Institutions (20220718110918001), and the Natural Science Foundation of Top Talent at SZTU (GDRC202131). The CMU team thanks the Student Travel Support for the ACM International Conference on Multimedia (ACM MM) and acknowledges the support of NSF CISE under grant number 1937998.